\documentclass[showpacs,aps,graphicx,twocolumn]{revtex4}%and
\usepackage{graphicx}
\begin{document}

\title{Two-step measurement of the concurrence for  hyperentangled state}

\author{ Yu-Bo Sheng,$^{1,2}$\footnote{Email address:
shengyb@njupt.edu.cn} and Lan Zhou$^{2,3}$ }
\address{$^1$ Institute of Signal Processing  Transmission, Nanjing
University of Posts and Telecommunications, Nanjing, 210003, China\\
 $^2$ Key Lab of Broadband Wireless Communication and Sensor Network
 Technology, Nanjing University of Posts and Telecommunications, Ministry of
 Education, Nanjing, 210003, China\\
 $^3$College of Mathematics \& Physics, Nanjing University of Posts and Telecommunications, Nanjing,
210003, China}

\date{\today } \begin{abstract}
We describe an efficient way for measuring the concurrence of the hyperentanglement. In this protocol, the hyperentangled state is encoded in both
polarization and momentum degrees of freedom. We show that the concurrences of both polarization and momentum entanglement can be conversed into the total success
probability of picking up the odd-parity state and can be measured directly.  This protocol requires the weak cross-Kerr nonlinearity to construct the quantum nondemolition measurement and does not resort to the sophisticated controlled-not gate operation. It is feasible in current experimental technology.
\end{abstract}

 \maketitle

\section{introduction}
Entanglement is the key element in quantum information processing \cite{book}. Nearly all quantum communication and computation protocols
require the entanglement. For example, in quantum  teleportation \cite{teleportation}, quantum key distribution \cite{Ekert91}, quantum state sharing \cite{QSS1,QSS2},
and other quantum communication protocols \cite{long,two-step}, they should require the entanglement to set up the channel.
In a quantum computation model, they also need to construct the entanglement \cite{computation1}. Before performing such protocols, they usually should know the
exact information of the entanglement. How to quantify entanglement  becomes an important and interesting topic in both
theory and experiment.
  In 1996, Bennett \emph{et al.} proposed the concept of the entanglement of formation to quantify entanglement \cite{concurrence1}. For a two-qubit pure state, the entanglement
  of formation can be exactly quantified by the concurrence. The concurrence is \cite{concurrence1,concurrence2,concurrence3}
  \begin{eqnarray}
  C=|\langle\Psi^{\ast}|\sigma_{y}\otimes\sigma_{y}|\Psi\rangle|.
  \end{eqnarray}
  If an arbitrary two-qubit pure state described as
  $|\varphi\rangle=\alpha|00\rangle+\beta|01\rangle+\gamma|10\rangle+\delta|11\rangle$, its concurrence is defined as
  $C(|\varphi\rangle)=2|\alpha\beta-\gamma\delta|$.  Here $|\alpha|^{2}+|\beta|^{2}+|\gamma|^{2}+|\delta|^{2}=1$. On the other hand, if the two-qubit state is simplified  to the partially entangled state $|\varphi'\rangle=\alpha|00\rangle+\beta|11\rangle$, we can easily obtain the concurrence as $C(|\varphi'\rangle)=2|\alpha\beta|$, with $|\alpha|^{2}+|\beta|^{2}=1$.

  In the early  work of Walborn \emph{et al.}, they have demonstrated the detection of the concurrence with linear optics \cite{concurrence4}.
  In 2007, Romero \emph{et al.}  described the way for measuring the concurrence of atomic-qubit pure state \cite{concurrence5}. In their protocol, they require the
  controlled-not (CNOT) gate between two atoms to complete the task. In 2008, the protocol for measuring the concurrence in a cavity QED system was proposed.
  They used  the atoms as the flying qubits to perform the measurement \cite{concurrence6}. The measurement for concurrence based on trapped ions were also proposed \cite{yangrc}. Recently, with the help of cross-Kerr nonlinearity, the group of Cao proposed two different methods for measuring the concurrence \cite{concurrence7,concurrence8}.

   Hyperentanglement, the simultaneous entangle
in more than one degree of freedom, which is widely studied in the recent years \cite{hyper1,hyper2,hyper3,hyperadd}.  The hyperentanglement can be used to  complete the Bell-state analysis \cite{hyperbell1,hyperbell2,hyperbell3,hyperbell4,hyperbell5,hyperbell6,hyperbell7,renbell,wangtj}, perform the entanglement purification \cite{Simon1,shengpra3,shengpra4,lixh,ren1}, and entanglement concentration \cite{ren2}.  It can also be used to extend the capacity of the channel in dense coding\cite{hyperbell6}.   Interestingly,  Walborn \emph{et al.} showed that the hyperentanglement can also be used to detect the concurrence of the polarization entanglement \cite{concurrence4}. In their protocol, they first produce the hyperentanglement in both polarization and momentum (spatial mode) degrees of freedom. Subsequently, they perform a CNOT gate
in one of the photon between the momentum and polarization degree of freedom. Finally, by detecting both the photons, they can measure the concurrence of the polarization entanglement.

Current works for detecting the concurrence all focus  on the entanglement in single degree of freedom. Though Walborn \emph{et al.} reported the detection of the concurrence with hyperentanglement, it only used to measure the concurrence for polarization entanglement. They did not provide a complete description for
hyperentanglement.   In this paper, we will describe an effect  way for measuring the concurrence of hyperentanglement. We show that the concurrence in each
degrees of freedom can be measured independently.  We resort to the weak cross-Kerr nonlinearity to construct the quantum nondemolition measurement. This paper is organized as follows: In Sec. II, we first provide an alternative definition for the concurrence of a hyperentangled state. In Sec. III, we explain our protocol with a simple example. We take the hyperentangled state encoded in the polarization and momentum entanglement as an example. Both the polarization and momentum entanglement are the pure partially entangled states. In Sec. IV, we will prove that the concurrence of arbitrary pure hyperentangled state can also be measured. In Sec. V, we will make a discussion and  conclusion.

 \section{Concurrence of the hyperentanglement}
 Hyperentanglement is the entanglement which simultaneous entangle in more than one degree of freedom. The hyperentanglements include  polarization-momentum entanglement,
polarization-time-bin entanglement, and polarization-
spatial modes-energy-time entanglement, and so on. Taking advantage of an enlarged Hilbert space, the hyperentangled state can be described as the product of $N$ Bell states as the form \cite{rmpentanglement,witness1}
 \begin{eqnarray}
 |\Upsilon\rangle=|\Theta_{1}\rangle\otimes|\Theta_{2}\rangle\cdots|\Theta_{N}\rangle.
 \end{eqnarray}
Here $N$ is the number of the degree of freedom. We denote the concurrence of $|\Theta_{i}\rangle$ as $C_{i} (i=1,2,\cdots N)$.
We can obtain the total concurrence of the hyperentangled state as
 \begin{eqnarray}
C_{hyper}=\sum_{i=1}^{N}C_{N}.
 \end{eqnarray}
 Form above definition, if all the $|\Theta_{i}\rangle$ are the maximally Bell states,  we can get $C_{hyper}=N$.

\section{Measuring the concurrence for partially hyperentangled state}
Before we start to explain this protocol, we first briefly introduce the key element, i. e. the cross-Kerr nonlinearity.
It can be used to perform the quantum nondemolition (QND) measurement, which has been widely used in quantum information processing, such as construction
of the CNOT gate \cite{QND1,lin1}, performing the Bell-state analysis\cite{QND2,hyperbell7}, entanglement purification and concentration \cite{shengpra,shengpra2,shengsingle,shengwstateconcentration}, and so on \cite{he1,he2,lin2,qubit1,qubit2,qubit3,zhangshou}. The Hamiltonian of a cross-Kerr nonlinear interaction is $H=\hbar\chi a^{\dagger}_{s}a_{s}a^{\dagger}_{p}a_{p}$ \cite{QND1}. Here the $a^{\dagger}_{s}$,$a_{s}$($a^{\dagger}_{p}$, $a_{p}$) are the creation and destruction operators of the signal (probe) mode. As shown in Fig. 1, We consider that a signal state $\mu|0\rangle+\nu|1\rangle$ is in the $a1$ spatial mode. This signal state combined with the coherent state
$|\alpha\rangle$ will couple with the cross-Kerr material. Here $|0\rangle$ and $|1\rangle$ are the photon number. The whole system will evolve as
 \begin{eqnarray}
 (\mu|0\rangle+\nu|1\rangle)|\alpha\rangle\rightarrow\mu|0\rangle|\alpha\rangle+\nu|1\rangle|\alpha e^{i\theta}\rangle.
 \end{eqnarray}
Here $\theta=\chi t$ with $t$ being the interaction time. It shows that the coherent state picks up a phase shift $\theta$ which is directly proportional to the
photon number of the signal state. Therefore, through the measurement of the phase of the coherent state, one can
obtain the information of the photon number of the signal state. It is so called the QND measurement.

We first describe the way for measuring the concurrence of the momentum (spatial mode) entanglement. Suppose that the hyperentangled state
can be described as follows
\begin{eqnarray}
|\psi\rangle&=&(\alpha_{1}|H_{a}\rangle|H_{b}\rangle+\beta_{1}|V_{a}\rangle|V_{b}\rangle)\nonumber\\
&\otimes&(\alpha_{2}|a1\rangle|b1\rangle+\beta_{2}|a2\rangle|b2\rangle).\label{hyper}
\end{eqnarray}

\begin{figure}[!h]%[tpb]
\begin{center}
\includegraphics[width=9cm,angle=0]{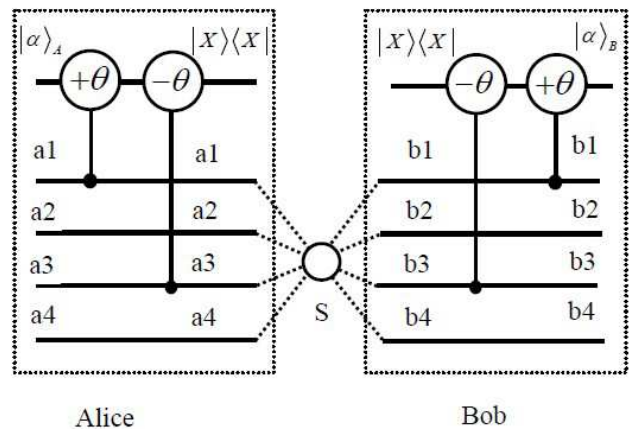}
\caption{Schematic of the measurement for the momentum entanglement. In each round, two pairs of hyperentangled states are emitted
by the source (S). It can make the parity-check for the momentum entanglement. }
\end{center}
\end{figure}

The two photons are distributed to Alice and Bob, respectively.
Here $|\alpha_{1}|^{2}+|\beta_{1}|^{2}=1$, and $|\alpha_{2}|^{2}+|\beta_{2}|^{2}=1$. $a1$, $b1$, $a2$ and $b2$ are the different spatial modes, as shown
in Fig. 1. The $|H\rangle$ is the horizontal polarization and $|V\rangle$  is the vertical polarization photon, respectively.
Such state can be generated by spontaneous parametric down-conversion source. As described in Ref. \cite{Simon1}, the pump
pulse of ultraviolet light passes through a $\beta$-barium borate
crystal (BBO). A correlated pair of photons will be generated
with the probability $p$ in the spatial modes $a1$ and $b1$.  Certainly, there is another
probability $p^{2}$ that generate two correlated pairs of photons in the spatial modes  $a1$ and $b1$.
The pulse can also be
reflected by the mirror and traverses the crystal a second time, producing
another correlated pair into the spatial modes $a2$ and $b2$ with the same
probability $p$.
If $p\ll1$, the $p^{2}$ can be omitted. In an ideal case, one can generate the hyperentanglement as shown in Eq. (\ref{hyper}).
Certainly, Ref. \cite{concurrence4} also provided another efficient way to generate the hyperentanglement.

As shown in Fig. 1, we choose two copies of hyperentangled states in each round. One is  $|\psi\rangle_{1}$ in the spatial modes $a1$, $b1$, $a2$ and $b2$, and the other
is $|\psi\rangle_{2}$  in the spatial modes $a3$, $b3$, $a4$ and $b4$, respectively.
The states $|\psi\rangle_{1}$ and $|\psi\rangle_{2}$ combined with two coherent states evolve as
\begin{eqnarray}
&&|\psi\rangle_{1}\otimes|\psi\rangle_{2}\otimes|\alpha\rangle_{A}\otimes|\alpha\rangle_{B}\nonumber\\
&=&(\alpha_{1}|H_{a}\rangle|H_{b}\rangle+\beta_{1}|V_{a}\rangle|V_{b}\rangle)\nonumber\\
&\otimes&(\alpha_{2}|a1\rangle|b1\rangle+\beta_{2}|a2\rangle|b2\rangle)
\otimes(\alpha_{1}|H_{a}\rangle|H_{b}\rangle+\beta_{1}|V_{a}\rangle|V_{b}\rangle)\nonumber\\
&\otimes&(\alpha_{2}|a3\rangle|b3\rangle+\beta_{2}|a4\rangle|b4\rangle)\otimes|\alpha\rangle_{A}\otimes|\alpha\rangle_{B}\nonumber\\
&\rightarrow&(\alpha^{2}_{2}|a1\rangle|a3\rangle|b1\rangle|b3\rangle|\alpha\rangle_{A}\otimes|\alpha\rangle_{B}\nonumber\\
&+&\alpha_{2}\beta_{2}|a1\rangle|a4\rangle|b1\rangle|b4\rangle|\alpha e^{i\theta}\rangle_{A}\otimes|\alpha e^{i\theta}\rangle_{B}\nonumber\\
&+&\alpha_{2}\beta_{2}|a2\rangle|a3\rangle|b2\rangle|b3\rangle|\alpha e^{-i\theta}\rangle_{A} \otimes|\alpha e^{-i\theta}\rangle_{B}\nonumber\\
&+&\beta^{2}_{2}|a2\rangle|a4\rangle|b2\rangle|b4\rangle|\alpha\rangle_{A}\otimes|\alpha\rangle_{B})\nonumber\\
&\otimes&(\alpha_{1}|H_{a}\rangle|H_{b}\rangle+\beta_{1}|V_{a}\rangle|V_{b}\rangle)\nonumber\\
&\otimes&(\alpha_{1}|H_{a}\rangle|H_{b}\rangle+\beta_{1}|V_{a}\rangle|V_{b}\rangle).\label{paritycheck1}
\end{eqnarray}

From Eq. (\ref{paritycheck1}), if both coherent states pick up the phase shift  $\theta$, the state becomes
\begin{eqnarray}
|\Phi\rangle_{1}&=&|a_{1}\rangle|a4\rangle|b1\rangle|b4\rangle\otimes(\alpha_{1}|H_{a}\rangle|H_{b}\rangle+\beta_{1}|V_{a}\rangle|V_{b}\rangle)\nonumber\\
&\otimes&(\alpha_{1}|H_{a}\rangle|H_{b}\rangle+\beta_{1}|V_{a}\rangle|V_{b}\rangle),\label{collapse1}
\end{eqnarray}
with the probability of $P_{m}=|\alpha_{2}\beta_{2}|^{2}$. Such measurement can be finished by a general homodyne-heterodyne measurement \cite{QND1}. The subscript $m$ means the momentum entanglement.

On the other hand, if both the coherent state  pick up the phase shift of $-\theta$, the state becomes
\begin{eqnarray}
|\Phi\rangle_{2}&=&|a2\rangle|a3\rangle|b2\rangle|b3\rangle \otimes(\alpha_{1}|H_{a}\rangle|H_{b}\rangle+\beta_{1}|V_{a}\rangle|V_{b}\rangle)\nonumber\\
&\otimes&(\alpha_{1}|H_{a}\rangle|H_{b}\rangle+\beta_{1}|V_{a}\rangle|V_{b}\rangle),\label{collapse2}
\end{eqnarray}
with the same probability of $|\alpha_{2}\beta_{2}|^{2}$.

Otherwise, if both coherent states pick up no phase shift, the state becomes
\begin{eqnarray}
|\Phi\rangle_{3}&=&(\alpha^{2}_{2}|a1\rangle|a3\rangle|b1\rangle|b3\rangle+\beta^{2}_{2}|a2\rangle|a4\rangle|b2\rangle|b4\rangle)\nonumber\\
&\otimes&(\alpha_{1}|H_{a}\rangle|H_{b}\rangle+\beta_{1}|V_{a}\rangle|V_{b}\rangle)\nonumber\\
&\otimes&(\alpha_{1}|H_{a}\rangle|H_{b}\rangle+\beta_{1}|V_{a}\rangle|V_{b}\rangle),\label{collapse3}
\end{eqnarray}
with the probability of $|\alpha^{4}|+|\beta|^{4}$.
From Eqs. (\ref{collapse1}) and (\ref{collapse2}), we can obtain the concurrence of the momentum entanglement
\begin{eqnarray}
C_{m}=2|\alpha_{2}\beta_{2}|=2\sqrt{P_{m}}.
\end{eqnarray}
Interestingly, from above description, it is shown that the polarization entanglement is not affected during the operation
on the momentum entanglement. Therefore, we can measure the concurrence of the polarization entanglement in the next round.
\begin{figure}[!h]%[tpb]
\begin{center}
\includegraphics[width=9cm,angle=0]{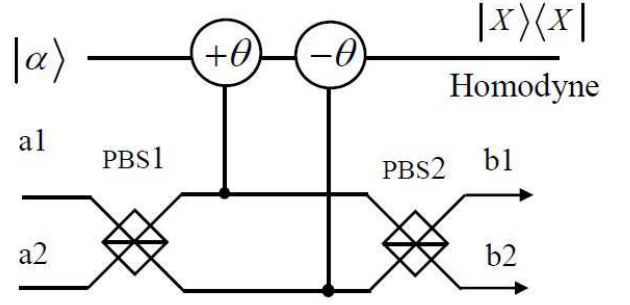}
\caption{Schematic drawing of the parity-check measurement for the polarization entanglement. The PBSs are the polarization beam splitters, which
can transmit the $|H\rangle$ polarization photon and reflect the $|V\rangle$ polarization photon.}
\end{center}
\end{figure}

For example, suppose that the initial state becomes $|\Phi\rangle_{1}$ after the measurement. From $|\Phi\rangle_{1}$, it can be rewritten as
\begin{eqnarray}
|\Phi_{1}\rangle&=&(\alpha_{1}|H_{a1}\rangle|H_{b1}\rangle+\beta_{1}|V_{a1}\rangle|V_{b1}\rangle)\nonumber\\
&\otimes&(\alpha_{1}|H_{a4}\rangle|H_{b4}\rangle+\beta_{1}|V_{a4}\rangle|V_{b4}\rangle)\nonumber\\
&=&\alpha^{2}_{1}|H_{a1}\rangle|H_{a4}\rangle|H_{b1}\rangle|H_{b4}\rangle+\beta^{2}_{1}|V_{a1}\rangle|V_{a4}\rangle|V_{b1}\rangle|V_{b4}\rangle\nonumber\\
&+&\alpha_{1}\beta_{1}(|H_{a1}\rangle|V_{a4}\rangle|H_{b1}\rangle|V_{b4}\rangle+|V_{a1}\rangle|H_{a4}\rangle|V_{b1}\rangle|V_{b4}\rangle.\label{collapse4}\nonumber\\
\end{eqnarray}
It means that  one photon pair is in the spatial mode $a1$ and $b1$, and the other is in the $a4$ and $b4$.
Then they let the photons in the $a1$ and  $a4$ spatial modes pass through the set up  shown in Fig. 2. In Fig. 2, the
two spatial modes are denoted as $a1$ and $a2$, respectively. The PBS is the polarization beam splitter, which can transmit
the $|H\rangle$ polarization photon and reflect the $|V\rangle$ polarization photon. Obviously, if the coherent state picks up the phase
shift $\pm \theta$, the state in Eq. (\ref{collapse4}) becomes
\begin{eqnarray}
|\Phi\rangle_{4}=\frac{1}{\sqrt{2}}(|H_{a1}\rangle|V_{a4}\rangle|H_{b1}\rangle|V_{b4}\rangle+|V_{a1}\rangle|H_{a4}\rangle|V_{b1}\rangle|V_{b4}\rangle),\nonumber\\
\end{eqnarray}
with the success probability of $P_{p}=2|\alpha_{1}\beta_{1}|^{2}$. In this way, we can obtain the concurrence of the polarization entanglement
\begin{eqnarray}
C_{p}=2|\alpha_{2}\beta_{2}|=\sqrt{P_{p}}.
\end{eqnarray}
Here the subscript $p$ means the polarization. Here we should point out that we should adopt the different way to measure the coherent state, which makes the
phase shift $\pm\theta$ undistinguished. This measurement can be achieved by choosing the local oscillator phase $\pi/2$ offset from the
 probe phase, which is called an $X$ quadrature measurement \cite{QND1}.
The total concurrence is
\begin{eqnarray}
C_{hyper}=C_{m}+C_{p}=2\sqrt{P_{m}}+\sqrt{P_{p}}.
\end{eqnarray}
So far, we have fully described our protocol with a simple example. The total protocol can be divided into two steps.  The first step is to measure the momentum entanglement and the second step is to
measure the polarization entanglement.  In our protocol, the concurrences have been transformed to the success probability of picking up the odd parity states, such as $|a1\rangle|a4\rangle$, $|b1\rangle|b4\rangle$ in spatial modes, and $|H\rangle|V\rangle$ and $|V\rangle|H\rangle$ in polarization entanglement, respectively. In order to complete the exact measurement of the  concurrence, we should perform the process many rounds and consume many photon pairs shown in Eq. (\ref{hyper}).
 This protocol can be realized on the fact that the hyperentangled states of the form of Eq. (\ref{hyper}) in two degrees of freedom can be operated independently.
It means that if we manipulate the momentum entanglement, we leave the polarization entanglement unchanged. Certainly, if we operate the polarization entanglement, the momentum entanglement does not change too. This advantage essentially provides us the effective way for performing this protocol. In order to measure the phase shift of the coherent state, they adopt two different measurement techniques. In the first step, they adopt the general homodyne-heterodyne measurement to pick up the $\theta$, while in the second step, they use the $X$ quadrature measurement to make the $\pm \theta$ undistinguished. Actually, in the first step, they can also make the $\pm \theta$ undistinguished, with the total success probability of $2|\alpha_{2}\beta_{2}|^{2}$. However, after performing this measurement, the system in  momentum degree of freedom is still entangled. Before measuring the polarization entanglement, they should add another QND to destroy the momentum entanglement. In our protocol, after the state picking up the phase shift  $\theta$, or $-\theta$, the spatial modes of the photons are essentially deterministic. They can start the second step directly.

\section{Measuring the concurrence of arbitrary hyperentangled state}
In above section, we have explained the method of measuring the concurrence for the hyperentangled state with four different coefficients.
Actually, this method can be extended to measure the concurrence of arbitrary hyperentangled state of the form
\begin{eqnarray}
|\phi\rangle&=&|\phi\rangle_{p}\otimes|\phi\rangle_{m}=(\alpha_{1}|H_{a}\rangle|H_{b}\rangle+\beta_{1}|V_{a}\rangle|V_{b}\rangle\nonumber\\
&+&\gamma_{1}|H_{a}\rangle|V_{b}\rangle+\delta_{1}|V_{a}\rangle|H_{b}\rangle)\nonumber\\
&\otimes&(\alpha_{2}|a1\rangle|b1\rangle+\beta_{2}|a2\rangle|b2\rangle\nonumber\\
&+&\gamma_{2}|a1\rangle|b2\rangle+\delta_{2}|a2\rangle|b1\rangle).\label{hyper2}
\end{eqnarray}
Here $|\alpha_{1}|^{2}+|\beta_{1}|^{2}+|\gamma_{1}|^{2}+|\delta_{1}|^{2}=1$, and $|\alpha_{2}|^{2}+|\beta_{2}|^{2}+|\gamma_{2}|^{2}+|\delta_{2}|^{2}=1$.
Following the same principle, we first describe the momentum entanglement of the form
 \begin{eqnarray}
 |\phi\rangle_{m}=\alpha_{2}|a1\rangle|b1\rangle+\beta_{2}|a2\rangle|b2\rangle
 +\gamma_{2}|a1\rangle|b2\rangle+\delta_{2}|a2\rangle|b1\rangle.\nonumber\\
\end{eqnarray}
\begin{figure}[!h]%[tpb]
\begin{center}
\includegraphics[width=8cm,angle=0]{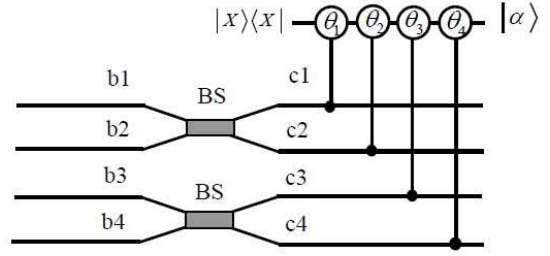}
\caption{Schematic drawing of measuring the momentum entanglement. In the last step, we should determine the spatial modes of the state to
perform further measurement. The BS is the 50:50 beam splitter.}
\end{center}
\end{figure}
As the entanglement in each degree of freedom can be operated independently,  we omit the polarization entanglement for simple in the following description.
The first step is similar to the previous section. From Fig. 1, they let the four photons pass through the two QNDs. If both the coherent states
pick up the phase shift $\theta$, the state becomes
\begin{eqnarray}
|\phi\rangle_{1}&=&\frac{\alpha_{2}\beta_{2}}{\sqrt{2(|\alpha_{2}\beta_{2}|^{2}+|\gamma_{2}\delta_{2}|^{2})}}(|a1\rangle|a4\rangle|b1\rangle|b4\rangle\nonumber\\
&+&|a2\rangle|a3\rangle|b2\rangle|b3\rangle\nonumber\\
&+&\frac{\gamma_{2}\delta_{2}}{\sqrt{2(|\alpha_{2}\beta_{2}|^{2}+|\gamma_{2}\delta_{2}|^{2})}}(|a1\rangle|a4\rangle|b2\rangle|b3\rangle\nonumber\\
&+&|a2\rangle|a3\rangle|b1\rangle|b4\rangle).
\end{eqnarray}
 The total success probability is $P_{1m}=2(|\alpha_{2}\beta_{2}|^{2}+|\gamma_{2}\delta_{2}|^{2})$.

Then they let the photons in the spatial modes $b1$, $b2$, $b3$ and $b4$ pass through the two beam splitters (BSs).
Two BSs will make
\begin{eqnarray}
|b1\rangle\rightarrow\frac{1}{\sqrt{2}}(|c1\rangle+|c2\rangle),\nonumber\\
|b2\rangle\rightarrow\frac{1}{\sqrt{2}}(|c1\rangle-|c2\rangle),\nonumber\\
|b3\rangle\rightarrow\frac{1}{\sqrt{2}}(|c3\rangle+|c4\rangle),\nonumber\\
|b4\rangle\rightarrow\frac{1}{\sqrt{2}}(|c3\rangle-|c4\rangle).
\end{eqnarray}
After passing through the BSs, the state $|\phi\rangle_{1}$ becomes
\begin{eqnarray}
|\phi\rangle'_{1}&=&\frac{\alpha_{2}\beta_{2}+\gamma_{2}\delta_{2}}{2\sqrt{2(|\alpha_{2}\beta_{2}|^{2}+|\gamma_{2}\delta_{2}|^{2})}}(|a1\rangle|a4\rangle+|a2\rangle|a3\rangle)\nonumber\\
&\otimes&(|c_{1}\rangle|c_{3}\rangle-|c_{2}\rangle|c_{4}\rangle)\nonumber\\
&+&\frac{\alpha_{2}\beta_{2}-\gamma_{2}\delta_{2}}{2\sqrt{2(|\alpha_{2}\beta_{2}|^{2}+|\gamma_{2}\delta_{2}|^{2})}}(|a1\rangle|a4\rangle-|a2\rangle|a3\rangle)\nonumber\\
&\otimes&(|c1\rangle|c4\rangle-|c2\rangle|c3\rangle).\label{collapse5}
\end{eqnarray}
From Eq. (\ref{collapse5}), if the phase shit is $\theta_{1}+\theta_{4}$,
the state in Eq. (\ref{collapse5}) will become
\begin{eqnarray}
|\phi\rangle''_{1}=\frac{1}{\sqrt{2}}(|a1\rangle|a4\rangle-|a2\rangle|a3\rangle)|c1\rangle|c4\rangle.
\end{eqnarray}
The success probability is $P_{2m}=\frac{|\alpha_{2}\beta_{2}-\gamma_{2}\delta_{2}|^{2}}{4(|\alpha_{2}\beta_{2}|^{2}+|\gamma_{2}\delta_{2}|^{2})}$.
 If the coherent state picks up the phase shift   $\theta_{2}+\theta_{3}$,
 the state in Eq. (\ref{collapse5}) will become
\begin{eqnarray}
|\phi\rangle'''_{1}=\frac{1}{\sqrt{2}}(|a1\rangle|a4\rangle-|a2\rangle|a3\rangle)|c2\rangle|c3\rangle,
\end{eqnarray}
 with the same probability.
 The total success probability of obtaining the state $|\phi\rangle''_{1}$ or  $|\phi\rangle'''_{1}$
 is
 \begin{eqnarray}
 P_{m}&=&P_{1m}P_{2m}=2(|\alpha_{2}\beta_{2}|^{2}+|\gamma_{2}\delta_{2}|^{2})\frac{|\alpha_{2}\beta_{2}-\gamma_{2}\delta_{2}|^{2}}{4|\alpha_{2}\beta_{2}|^{2}+|\gamma_{2}\delta_{2}|^{2}}\nonumber\\
 &=&\frac{1}{2}|\alpha_{2}\beta_{2}-\gamma_{2}\delta_{2}|^{2}.
 \end{eqnarray}
 Therefore, we can obtain the concurrence
  \begin{eqnarray}
  C( |\phi\rangle_{m})=2|\alpha_{2}\beta_{2}-\gamma_{2}\delta_{2}|=2\sqrt{2P_{m}}.
  \end{eqnarray}
  In the next step, we  will describe the measurement of the concurrence of the polarization entanglement.
  However, before they start the second step of the protocol, they should first decide the spatial mode of the whole system.
 We take the state $|\phi\rangle''_{1}$ as an example.
 If we consider the polarization entanglement, the whole state should be rewritten as
 \begin{eqnarray}
|\phi\rangle_{1}&=&(\alpha_{1}|H_{a}\rangle|H_{c}\rangle+\beta_{1}|V_{a}\rangle|V_{c}\rangle\nonumber\\
&+&\gamma_{1}|H_{a}\rangle|V_{c}\rangle+\delta_{1}|V_{a}\rangle|H_{c}\rangle)\nonumber\\
&\otimes&(\alpha_{1}|H_{a}\rangle|H_{c}\rangle+\beta_{1}|V_{a}\rangle|V_{c}\rangle\nonumber\\
&+&\gamma_{1}|H_{a}\rangle|V_{c}\rangle+\delta_{1}|V_{a}\rangle|H_{c}\rangle)\nonumber\\
&\otimes&\frac{1}{\sqrt{2}}(|a1\rangle|a4\rangle-|a2\rangle|a3\rangle)|c1\rangle|c4\rangle.\label{hyper3}
\end{eqnarray}
 From Eq. (\ref{hyper3}), it is shown that the two photons in Bob's location are  deterministic in  $c1$ and $c4$,
 while the photons in the Alice's location are still entangled in the spatial modes. Therefore, they let the photons in the $a1$, $a2$, $a3$
 and $a4$ pass through
 the set up shown in Fig. 3 by removing two BSs. If the phase shift is $\theta_{1}+\theta_{4}$, $|\phi\rangle_{1}$ becomes
  \begin{eqnarray}
  |\phi\rangle'_{1}&=&(\alpha_{1}|H_{a1}\rangle|H_{c1}\rangle+\beta_{1}|V_{a1}\rangle|V_{c1}\rangle\nonumber\\
&+&\gamma_{1}|H_{a1}\rangle|V_{c1}\rangle+\delta_{1}|V_{a1}\rangle|H_{c1}\rangle)\nonumber\\
&\otimes&(\alpha_{1}|H_{a4}\rangle|H_{c4}\rangle+\beta_{1}|V_{a4}\rangle|V_{c4}\rangle\nonumber\\
&+&\gamma_{1}|H_{a4}\rangle|V_{c4}\rangle+\delta_{1}|V_{a4}\rangle|H_{c4}\rangle).\label{polarzation1}
\end{eqnarray}
Otherwise, if the phase shift is  $\theta_{2}+\theta_{3}$,
$|\phi\rangle_{1}$ becomes
 \begin{eqnarray}
  |\phi\rangle''_{1}&=&(\alpha_{1}|H_{a2}\rangle|H_{c1}\rangle+\beta_{1}|V_{a2}\rangle|V_{c1}\rangle\nonumber\\
&+&\gamma_{1}|H_{a2}\rangle|V_{c1}\rangle+\delta_{1}|V_{a2}\rangle|H_{c1}\rangle)\nonumber\\
&\otimes&(\alpha_{1}|H_{a3}\rangle|H_{c4}\rangle+\beta_{1}|V_{a3}\rangle|V_{c4}\rangle\nonumber\\
&+&\gamma_{1}|H_{a3}\rangle|V_{c4}\rangle+\delta_{1}|V_{a3}\rangle|H_{c4}\rangle).\label{polarzation1}
\end{eqnarray}
After the measurement, the momentum entanglement is disappeared, and the spatial mode of each photon is deterministic. In this way,
they can start the detection of the concurrence of the polarization entanglement. The way of measuring the polarization entanglement is similar to
the Ref.\cite{concurrence7}. Suppose that both Alice and Bob own the QND as shown in Fig. 2. They first let the four photons pass through the QND, and pick up the odd parity
state. Therefore,  $|\phi\rangle'_{1}$ will become
\begin{eqnarray}
 |\phi\rangle'_{2}&=&\frac{\alpha_{1}\beta_{1}}{\sqrt{2(|\alpha_{1}\beta_{1}|^{2}+|\gamma_{1}\delta_{1}|^{2})}}(|H_{a1}\rangle|V_{a4}\rangle|H_{c1}\rangle|V_{c4}\rangle\nonumber\\
  &+&|V_{a1}\rangle|H_{a4}\rangle|V_{c1}\rangle|H_{c4}\rangle\nonumber\\
  &+&\frac{\gamma_{1}\delta_{1}}{\sqrt{2(|\alpha_{1}\beta_{1}|^{2}+|\gamma_{1}\delta_{1}|^{2})}}(|H_{a1}\rangle|V_{a4}\rangle|V_{c1}\rangle|H_{c4}\rangle\nonumber\\
  &+&|V_{a1}\rangle|H_{a4}\rangle|H_{c1}\rangle|V_{c4}\rangle),
\end{eqnarray}
with the success probability $P_{1p}=2(|\alpha_{1}\beta_{1}|^{2}+|\gamma_{1}\delta_{1}|^{2})$. In the second round, they first perform the Hadamard
operation on the photons in $a1$ and $a4$ mode.
The Hadamard operation can be implemented with the quarter wave plate (QWP), and makes
\begin{eqnarray}
|H\rangle\rightarrow\frac{1}{\sqrt{2}}(|H\rangle+|V\rangle),
|V\rangle\rightarrow\frac{1}{\sqrt{2}}(|H\rangle-|V\rangle).
\end{eqnarray}
After performing the Hadamard operations, the state $|\phi\rangle'_{2}$ will become
\begin{eqnarray}
 |\phi\rangle'_{3}&=&\frac{\alpha_{1}\beta_{1}+\gamma_{1}\delta_{1}}{2\sqrt{2(|\alpha_{1}\beta_{1}|^{2}+|\gamma_{1}\delta_{1}|^{2})}}(|H_{a1}\rangle|H_{a4}\rangle-|V_{a1}\rangle|V_{a4}\rangle)\nonumber\\
&\otimes&(|H_{c1}\rangle|V_{c4}\rangle+|V_{c1}\rangle|H_{c4}\rangle)\nonumber\\
&+&\frac{\alpha_{1}\beta_{1}-\gamma_{1}\delta_{1}}{2\sqrt{2(|\alpha_{1}\beta_{1}|^{2}+|\gamma_{1}\delta_{1}|^{2})}}(|H_{a1}\rangle|V_{a4}\rangle-|V_{a1}\rangle|H_{a4}\rangle)\nonumber\\
&\otimes&(|H_{c1}\rangle|V_{c4}\rangle-|V_{c1}\rangle|H_{c4}\rangle).\label{collapse6}
\end{eqnarray}
Finally,  they let the photons in $a1$ and $a4$ modes pass through the QND in Fig. 3 again,
and pick  up the odd parity state in a second time. It will make the state
 $|\phi\rangle'_{3}$ become
 \begin{eqnarray}
  |\phi\rangle'_{4}&=&\frac{1}{2}(|H_{a1}\rangle|V_{a4}\rangle-|V_{a1}\rangle|H_{a4}\rangle)\nonumber\\
&\otimes&(|H_{c1}\rangle|V_{c4}\rangle-|V_{c1}\rangle|H_{c4}\rangle),
 \end{eqnarray}
 with the success probability $P_{2p}=\frac{|\alpha_{2}\beta_{2}-\gamma_{2}\delta_{2}|^{2}}{2(|\alpha_{2}\beta_{2}|^{2}+|\gamma_{2}\delta_{2}|^{2})}$.
 Therefore, the total success probability
 \begin{eqnarray}
 P_{p}=P_{1p}P_{2p}=|\alpha_{1}\beta_{1}-\gamma_{1}\delta_{1}|^{2}.
  \end{eqnarray}
  We can obtain the concurrence
   \begin{eqnarray}
  C( |\phi\rangle_{p})=|\alpha_{2}\beta_{2}-\gamma_{2}\delta_{2}|=2\sqrt{P_{p}},
  \end{eqnarray}
  and the total concurrence
   \begin{eqnarray}
  C(|\phi\rangle)=C(|\phi\rangle_{m})+C(|\phi\rangle_{p})=2\sqrt{2P_{m}}+2\sqrt{P_{p}}.
  \end{eqnarray}

\section{discussion and conclusion}
So far, we have completely described our protocol. We first described the method for measuring the concurrence of the partially
hyperentangled state. Subsequently, we provided the way of measuring the arbitrary pure hyperentangled  state. This protocol can be
divided into two steps. In the first step, we perform the measurement for the momentum entanglement. In the second step, we describe the
measurement of the polarization entanglement.  It depends on the  distinct
feature  that the different degrees
of freedom are relatively independent. It essentially
ensures that one can manipulate each degree of freedom
independently. In the practical operation,  momentum entanglement should be measured first, for once the spatial modes
of the photon is deterministic, one can easily perform the further measurement for polarization entanglement. During the whole procedure,
we mainly explain the polarization entanglement measurement after successfully performing the momentum entanglement. Actually, the measurement
step does not affect the second one. Even if the measurement of the momentum entanglement is a failure, we can also perform the second step, after
determining the spatial mode by adding another QNDs. In this manner, we can improve the practical efficiency.

Usually, in the early works
of the generation of the hyperentanglement, the Bell inequalities are adopted to characterize the quality of the  hyperentanglement. Vallone \emph{ et al.} also introduced the hyperentanglement witness to detect a two-particle state is hyperentangled \cite{witness2}. Unfortunately, as pointed out by Walborn \emph{et al.},
the Bell inequalities and entanglement witness cannot provide satisfactory results in general, because they disclose the entanglement of some quantum states but fail for other states. Meanwhile,  they are fundamentally different from entanglement measurement that, by definition , quantify the amount of entanglement in any state. With the help of hyperentanglement, they have successfully determinate the polarization entanglement. However, in their experiment, the momentum entanglement acts as the auxiliary resources to perform the CNOT gate. Their protocol cannot describe the fully concurrence for hyperentanglement including both momentum entanglement and polarization entanglement. On the other hand, an direct method for detecting the entanglement would be the quantum state tomographic reconstruction \cite{tomography}. It requires 15 parameters to reconstruct a two qubit state. If we consider a two-qubit hyperentangled state in an enlarged Hilbert space,
it requires 255 parameters, which makes this method extremely complicated.

Finally, let us briefly discuss the QND, which plays the important role in this protocol. During the measurement, we should pick up the phase shift $\theta$ using the a general homodyne-heterodyne measurement, or pick up the phase shift $\pm\theta$ to make the $\pm\theta$ undistinguishable, using the $X$ quadrature measurement. The requirement for this technique is $\alpha\theta>1$, where  $\alpha$ is
the amplitude of the coherent state. Even with the weak nonlinearity, this requirement can be satisfied by choosing
large amplitude of the coherent state. Certainly, the largest natural cross-Kerr
nonlinearities are extremely
weak ($\chi^{3}\approx10^{-22}m^{2}V^{-2}$) \cite{kok1}, which causes some controversy for its application \cite{Gea,Shapiro1,Shapiro2}.
Fortunately, Xing \emph{et al.} showed that the weak measurement can be applied to amplify the phase shift to the observable value \cite{weak_meaurement}.
Current experiment showed that the "giant" cross-Kerr effect with phase shift of 20 degrees per photon has been observed \cite{gaint}.

In conclusion, we have described an effective way of measuring the concurrence for hyperentangled state. We first provide an alternative definition of the concurrence for the hyperentangled state. We show that each concurrence in different degrees of freedom can be measured independently.
 We do not require the sophisticated CNOT gate operation and resort to the feasible weak cross-Kerr nonlinearity to perform the parity-check measurement.
 This way of characterize the hyperentanglement may extremely useful in current quantum information processing.
\section*{ACKNOWLEDGEMENTS}
This work is supported by the National Natural Science Foundation of
China under Grant Nos. 11104159 and 11347110.  and the Project
Funded by the Priority Academic Program Development of Jiangsu
Higher Education Institutions.

  \end{document}